\documentclass[aps,prl,twocolumn,groupedaddress]{revtex4-1}
\usepackage{amsmath}
\usepackage{graphicx}

\begin{document}

\title{On spatial non-homogeneity in iron pnictides:
formation of stripes}
\author{Lev P. Gor'kov}
\affiliation{NHMFL, Florida State University, 1800 E Paul Dirac
Dr., Tallahassee FL 32310, USA} \affiliation{L.D.Landau Institute
for Theoretical Physics of the RAS, Chernogolovka 142432, RUSSIA }
\author{Gregory B. Teitel'baum}
\email{grteit@kfti.knc.ru}
\affiliation{E.K.Zavoiskii Institute for Technical Physics of the RAS, Kazan 420029,
RUSSIA }
\date{\today}

\begin{abstract}
The heterogeneous coexistence of antiferromagnetism (SDW)
and superconductivity on a mesoscopic scale was observed
in iron-pnictides in many recent experiments. We suggest
and discuss the scenario in which the heterogeneity is
caused by formation of domain
walls inherent to the SDW state of pnictides at a proper
doping or under applied pressure. Superconductivity would
emerge from the modulated SDW structure. The phenomenon
is akin to the FFLO-phase in superconductors.
\end{abstract}

\maketitle

The new vast class of layered HTSC materials, iron pnictides
\cite{Hosono1}, manifests diverse and uncommon magnetic,
structural and superconducting properties \cite{Physica C}. Below
we address competition between magnetism (SDW) and
superconductivity (SC) that take place near the so-called "quantum
critical point" (QCP) on the pnictides' phase diagram tuned by
either doping, or pressure. In experiments \cite{Park, Kitagawa}
the competition assumes a form of the heterogeneous phase
coexistence. We demonstrate that the heterogeneity can be due to
the new SDW phase, where the staggered magnetization is modulated
by the emergent periodic stripes, the lattice of solitons. In this
vicinity SC arises on the background of the Soliton phase (SP).
Density of states (DOS) in SP has the same order of magnitude as
in the paramagnetic (PM) phase.

Currently, the consensus is that the weak
coupling nesting model \cite{Singh} correctly describes the most
typical peculiarities of the pnictides', at least
qualitatively.

We confine ourselves to the picture of \emph{only two}
Fermi-surfaces (FS): the one for holes at the $\Gamma$-point,
$(0,0)$ -point and the other, for electrons, at the M-point,
($\pi$, $\pi$), in the folded BZ. The model is in broad
use in the literature (see, for instance, in \cite{Vavilov}) and
among other things reproduces the overall view of the phase
diagram as function of doping \cite{Vavilov}, the interplay and
the sequence of magnetic and structural transitions in the undoped
compounds \cite{BarGor1}. In particular,  by way of changing the
degree of nesting the model provides the built-in mechanism for
the competition between magnetism and superconductivity at doping
or under applied pressure.

Among many aspects of the original nesting model \cite{KelKop}
that were recently investigated afresh in numerous theoretical
papers, there exists one interesting feature that
deserves more attention. The phenomenon consists in appearance of
a spatially non-uniform SDW state in pnictides under rather
general conditions near QCP.

Our attention to such a possibility was attracted by experiments
\cite{Park, Kitagawa}. In \cite{Kitagawa} a spontaneous spatial
hybrid SDW/SC structure (of few nm) was reported in the
SrFe$_2$As$_2$-crystal under pressure. In \cite{Park} the
heterogeneous phase coexistence was observed in single underdoped
(Ba,K)Fe$_2$As$_2$ crystal both above and below SC $T_c$. The
spatial scale in \cite{Park} amounted to 65 nm.

In the scenario under investigation the tendency to form the
periodic domain structure is inherent into the SDW nesting
mechanism itself. Therefore, onset of the modulated SDW phase may
occur at temperatures even above SC $T_c$. Precisely such behavior
was found in the quasi-one-dimensional Bechgaard salt,
(TMTSF)$_2$PF$_6$, \cite{PF6new}. The modulated SDW (at a fixed
pressure, $P<P_{cr}$, i.e., before QCP) was seen above and below
the transition into the new SC state and interpreted in terms of
the new ground state: SP \cite{GorGrig}.

First, as it seems, modulated CDW or SDW phases in three
dimensions were discussed in \cite{Rice, GorMnaz}. In
\cite{GorMnaz} the authors investigated the model \cite{KelKop} of
the two anisotropic Fermi surfaces with the shapes deviating from
the perfect nesting. Consequently, the energy spectrum for
electron and holes was chosen as $H_{e,h}=\pm v_{F1,2}(t-\eta
_{1,2}(\overrightarrow{p}))$, correspondingly. Deviations from the
ideal nesting $\eta _{1}(\overrightarrow{p})$ and $\eta _{2}(%
\overrightarrow{p})$ describe both the anisotropy and the doping.

Analysis \cite{GorMnaz} had shown that the instability line with
respect to transition into a \emph{commensurate} CDW (or SDW)
phase in the ($T,\eta $) -plane possesses the  reentrant
character, similar to a superconductor placed into the exchange field, $I\sigma
_{z}$ \cite{GorRus, LO}. On the side of large (\textquotedblleft
antinesting\textquotedblright
) terms, $\eta _{1}(\overrightarrow{p})$ and $\eta _{2}(\overrightarrow{p})$%
, the onset of SDW first occurs at an incommensurate (IC) vector
$\overrightarrow{Q}=\overrightarrow{Q}_{0}+\overrightarrow{q}$
where $\overrightarrow{Q}_{0}(\pi $,$\pi $$)$ is the commensurate
vector at zero $\eta _{1}(\overrightarrow{p}),$ $\eta
_{2}(\overrightarrow{p})$. At $T=0$ the $\vec{q}$-vector is
related to the magnitude and the anisotropy combined, as given by
the \textquotedblleft
antinesting\textquotedblright\ factor, $\left\vert \eta _{1}(\overrightarrow{%
p})-\eta _{2}(\overrightarrow{p})\right\vert $ \cite{GorMnaz}:
\begin{equation}
\langle \ln \left\{ \left\vert (\eta _{1}-\eta _{2})^{2}-(\cos \varphi
)^{2}q^{2}\right\vert (\frac{2v_{F1}v_{F2}\gamma}{\pi T_{c0}(v_{F1}+v_{F2})}%
)^{2}\right\} \rangle =0  \label{eq5}
\end{equation}%
(Average, $<\ldots >$, is taken along FS).

Appearance of an IC SDW on the phase diagram of pnictides was
discussed earlier (See for instance \cite{Vavilov, Tesanovic}).
However, we emphasize the profound difference between IC SDW and
the inhomogeneous SDW state, where electronic degrees of freedom
inside domain walls (DW) are not gapped which leads to a finite
and spatially inhomogeneous DOS (see below).

On the ($T,\eta $)-phase diagram in Fig.1 the instability line
(dashed) for the commensurate SDW ($\vec{q}=0$) displays
reentrance below the tri-critical point ($T_{tcr},\eta_{tcr}$);
the dotted line that goes down from the same point is the would-be
line of the 1st order transition between the commensurate SDW and
the paramagnetic phase. The periodic SP supersedes this
transition. The new phase  extends to the dash-dotted line  on the
right. The position of the QCP \emph{inside} the SC dome is shown
tentatively.
\begin{figure}[tbp]
\centering \includegraphics[height=6cm]{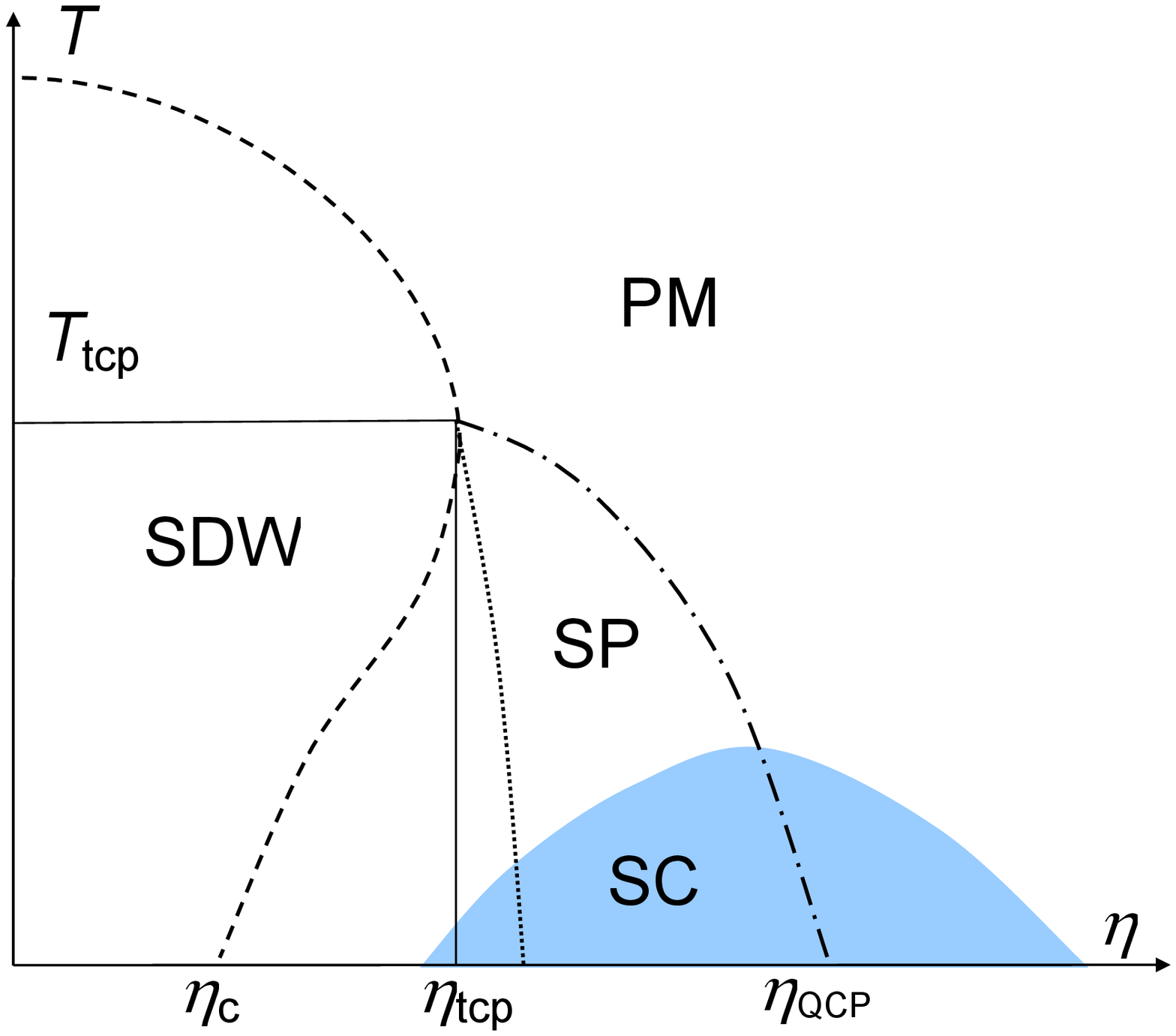} %\lable{fig:1}
\caption{The $(T, \eta)$ phase diagram ($\eta$ - tuning
parameter): the instability line (dashed) for commensurate SDW
shows reentrance below the tri-critical point
($T_{tcr},\eta_{tcr}$); SP starts before the putative 1-st order
transition line (dotted) and extends to the PM phase boundary (dash-dotted).
The QCP is covered by the SC dome. See text for the details.}
\end{figure}

We derive
the system of the Eilenberger-type equations describing the
inhomogeneous SP state. With the view to demonstrate the new
phenomenon, we accept the model \cite{GorMnaz} assuming a
density-density interaction between the \emph{e-} and \emph{h}-pockets.
The analytical solution for SP is not available. In particular,
if the ideal nesting is
broken by \textit{doping only}, the substitution,
$v_{F}$$\eta_{1}(\overrightarrow{p})=-v_{F}\eta_{2}(\overrightarrow{p}%
)=const\Rightarrow I$, leads to equations identical to the ones
for FFLO problem \cite{LO, FF}, solved numerically in
\cite{BurkRain}. Same methods apply to the equations for Fe
pnictides in the general case. Later on in the paper, we
investigate emergence of domains from the homogeneous SDW/CDW
state at the increasing strength of
the parameters, $\eta _{1}(\overrightarrow{p}),$ $\eta _{2}(\overrightarrow{p%
})$, say, by pressure.

Derivation of the Eilenberger-style equations for the
itinerant 3D or 2D SDW/CDW case
is similar to their original derivation \cite{Eilenb} from the
Gor'kov equations in superconductors and is described below only
briefly. (Similar equations were used for 1D physics in
\cite{ArtVolk}).

Introduce the matrix Green function
\begin{equation}
\hat{G}(x,x^{\prime })=\left(
\begin{array}{cc}
G_{11} & G_{12} \\
-G_{21} & -G_{22}%
\end{array}%
\right).  \label{eq6}
\end{equation}%
Here
$G_{ik}(x,x^{\prime })=\left\langle T(\overline{\psi }_{i}(x)\overline{\psi }%
_{i}^{+}(x^{\prime })\right\rangle$, $x=(\vec{r},\tau )$, and the
non-diagonal indices, 12, 21,  belong to the off-diagonal Green
functions, non-zero in the presence of
SDW/CDW-order. The two FS's are connected by the structure vector $%
\overrightarrow{Q}$. To save on the notations, we consider the CDW
case. The only difference is in the spin matrix,
($\vec{\sigma}\cdot\vec{l})$, present as a factor, in the
definitions of the non-diagonal SDW components, $G_{i\neq
k}(x,x^{\prime })$ ($\vec{l}$ stands for the staggered
magnetization direction).

We then write down the first pair of equation:
\begin{equation}
\begin{aligned}
(\partial /\partial \tau +\hat{H}_{e})G_{11}(x,x^{\prime })+\Delta _{\vec{Q}%
}(\,\/\vec{r}\,)G_{21}(x,x^{\prime })&=1\\
(\partial /\partial \tau +\hat{H}_{h})G_{21}(x,x^{\prime })+\Delta _{-\vec{Q}%
}^{\ast }(\,\/\vec{r}\,)G_{11}(x,x^{\prime })&=1
\label{eq9}
\end{aligned}
\end{equation}
Here:
\begin{equation}
\Delta _{\vec{Q}}(\vec{r})\equiv\lambda
G_{21}(\vec{r};\tau_{1}=\tau_{2});\text{ }\Delta
_{-\vec{Q}}^{\ast }(\vec{r})\equiv\lambda
G_{12}(\vec{r};\tau_{1}=\tau_{2})  \label{eq10}
\end{equation}
( $\lambda $ is proportional to the interaction constant).

To account for the spatial dependence in (\ref{eq10}), we write for free
electrons and holes:
\begin{equation}
\hat{H}_{e,h}=\mp ( \nabla ^{2}/2m_{1,2}+\mu )\mp v_{F1,2}\cdot\eta
_{1,2}(\overrightarrow{p}) \label{eq11}
\end{equation}
The \textquotedblleft gap\textquotedblright\ parameters (\ref{eq10}) couple
the electron and the hole FS's.
Following \cite{Eilenb}, introduce the operator in the L.H.S. of
Eqs.(\ref{eq9}):
\begin{equation}
\check{G}_{L}^{-1}=\left(
\begin{array}{cc}
\partial /\partial \tau +\hat{H}_{e} & -\Delta _{\vec{Q}}(\,\/\vec{r}\,) \\
\Delta _{-\vec{Q}}^{\ast }(\,\/\vec{r}\,) & -\partial /\partial \tau -\hat{H}%
_{h}%
\end{array}%
\right)  \label{eq13}
\end{equation}
so that $\check{G}_{L}^{-1}\times \hat{G}(x,x^{\prime })=\delta
(x-x^{\prime })$.  The operator that acts on the matrix (\ref{eq6}) from
the right side: $\hat{G}(x,x^{\prime })\times
\check{G}_{R}^{-1}=\delta (x-x^{\prime })$ is obtained by changing
$\partial /\partial \tau \rightarrow -\partial /\partial \tau$.

Substitute $\partial /\partial \tau \Longrightarrow -i\omega _{n}$
in $\check{G}_{L}^{-1}$ and $\partial /\partial \tau
\Longrightarrow +i\omega _{n}$ in $\check{G}_{R}^{-1}$ and rewrite
equations for the frequency Fourier components. The
Green functions are not diagonal in the momentum
representation:
\begin{equation}
\hat{G}(x,x^{\prime })\Longrightarrow \hat{G}(\omega _{n};\vec{r},\vec{r}%
\,^{\prime })\Longrightarrow \hat{G}(\omega _{n};\vec{p},\vec{p}-\vec{k})
\label{eq17}
\end{equation}

The essence of the \textquotedblleft
quasiclassical\textquotedblright\ method \cite{Eilenb, LOnew} in
the theory of superconductivity is in integrating out the energy
variable $\xi _{\vec{p}}=v_{F}t\equiv v_{F}(p-p_{F})$ thus rewriting equations in terms of new functions:
\begin{equation}
\begin{aligned}
\int \frac{d\xi _{\vec{p}}}{i\pi }\hat{G}(\omega _{n};\vec{p},\vec{p}-\vec{k}%
)\Longrightarrow \hat{g}(\omega _{n};\vec{p}_{F},\vec{k})\equiv\\ \equiv\left(
\begin{array}{cc}
  g_{11}(\omega _{n};\vec{p}_{F},\vec{k}) & g_{12}(\omega _{n};\vec{p}_{F},%
\vec{k}) \\
  -g_{21}(\omega _{n};\vec{p}_{F},\vec{k}) & -g_{22}(\omega _{n};\vec{p}_{F},%
\vec{k}) \\
\end{array}\right)
\label{10a}
\end{aligned}
\end{equation}

In a superconductor, the characteristic scale for $T_{c}$'s is
usually much less than the energy Fermi: $T_{c}\ll E_{F}$. With
the characteristic scale for the magnetic transition,
$T_{SDW}\sim$100-200 K and the Fermi pockets energy scale
$\sim$ 0.2 eV  \cite{Physica C}, the quasiclassical approach is
expected to work well enough for pnictides as well.

The equations for the integrated Green function in \cite{Eilenb,
LOnew} emerge from commuting the matrix equations for the Green
functions:   $\check{G}_{L}^{-1}\times \hat{G} - \hat{G}\times
\check{G}_{R}^{-1}=0$  from whence the variable $\xi _{\vec{p}}$
drops out. Omitting the arguments of $\hat{g}(\omega
_{n};\vec{p}_{F},\vec{k})$ and denoting
$\delta_{\vec{p}_{F}}$=$v_{F}[\eta
_{1}(\overrightarrow{p}_{F})-\eta _{2}(\overrightarrow{p}_{F})]$,
the resulting equations acquire the form:
\begin{equation}
\begin{aligned}
-i(\vec{v}_{F}\cdot \vec{\nabla})g_{11}+\Delta _{\vec{Q}}(\,\/\vec{r}\,)g_{21}
-\Delta _{-\vec{Q}}^{\ast }(\,\/\vec{r}\,)g_{12}&=0\ \ \ \ \ \\
\{-i(\vec{v}_{F}\cdot \vec{\nabla})-2i\omega
_{n}-\delta_{\vec{p}_{F}}\}g_{12}+2\Delta
_{\vec{Q}}(\,\/\vec{r}\,)g_{11}&=0\ \ \ \ \ \\
\{+i(\vec{v}_{F}\cdot \vec{\nabla})-2i\omega _{n}-\delta_{\vec{p}_{F}}\}g_{21}
+2\Delta _{-\vec{Q}}^{\ast }(\,\/\vec{r}%
\,)g_{11}&=0\
\label{eq19}
\end{aligned}
\end{equation}
As in \cite{Eilenb}, $g_{11}=g_{22}$ and the normalization
condition is $g_{11}^{2}-g_{12}g_{21}=1$. The self-consistency
Eqs.(\ref{eq10}) now is:
\begin{equation}
\Delta _{\vec{Q}}(\,\/\vec{r}\,)=\lambda T\int dS_{\overrightarrow{p}_{F}}%
\underset{n}{\sum }g_{21}(\omega _{n};\vec{p}_{F},\vec{k})  \label{eq22}
\end{equation}

Eqs.(\ref{eq19}, \ref{eq22}) form the closed system. (One drawback of
the "quasiclassical" method is that at the derivation one needs
$v_{F1}=v_{F2}$, $m_1=m_2$).  In Fe pnictides two
FS are connected by the commensurate vector
$\overrightarrow{Q}_{0} (\pi $,$\pi $$)$. Therefore $\Delta(r)$
is real and we omit the index $\vec{Q}$ below.

Parameters $\eta$ depend on the specific material. Therefore we
stay below only on the principial properties of the SP. Near the
right boundary of the area on the phase diagram in Fig.1 occupied
by SP, the amplitude of the SDW- parameter, $\Delta$, is small and
the periodic solution of the form $\Delta (r)=\Delta \cos
(\vec{q}\,\,\vec{r})$ can be constructed by perturbations in
Eqs.(\ref{eq9}, \ref{eq10}) \cite{LO, GorMnaz}. With the
\textit{decrease} of the $\eta $-parameters and entering the
developed SP, the period gradually increases; below a threshold at
a critical value of the $\eta $-parameters, the system enters the
homogeneous SDW phase. Slightly above the threshold the structure
consists of the almost isolated domains.

It presents the decided interest to consider formation of one
single DW at the threshold as it sheds more light on peculiarities
of SP. Assume the dependence of $\Delta (x)$ on one spatial
variable. Return to Eqs.(\ref{eq9}) and address the problem of the
eigen values and the two-component eigen functions $(u,v)$ for the
operator (\ref{eq11})  $(\partial /\partial \tau \Longrightarrow
-E)$:
\begin{equation}
\begin{aligned}
(-iv_{F,x}\partial /\partial x-v_{F}\eta _{1}(\vec{p}%
_{F}))u+\Delta(x)v &=E(v_{F,x},\vec{p}_F)u \\
(+iv_{F,x}\partial /\partial x+v_{F}\eta _{2}(\vec{p}%
_{F}))v+\Delta(x)u &=E(v_{F,x},\vec{p}_F)v
\label{eq24}
\end{aligned}
\end{equation}%

The substitution $u,v \Longrightarrow (\bar{u},\bar{v})\times \exp
(i[\frac{v_F}{2v_{F,x}}(\eta_1+\eta_2)]x)$   transforms (\ref{eq24}) into:
\begin{equation}
\begin{aligned}
-iv_{F,x}\partial \bar{u}/\partial x+\Delta(\,\/x\,)\bar{v} &=%
\tilde{E}(v_{F,x})\bar{u} \\
+iv_{F,x}\partial \bar{v}/\partial x+\Delta(\,\/x\,)\bar{%
u} &=\tilde{E}(v_{F,x})\bar{v}
\label{eq26}
\end{aligned}
\end{equation}
where
$\tilde{E}(v_{F,x})=E(v_{F,x},\vec{p}_F)+\delta_{\vec{p}_{F}}/2$.

Consider the energy spectrum for system (\ref{eq26})
in the presence of a rarified periodic array of DW.
Near the isolated wall choose the "gap", $\Delta(x)$, in the form:
\begin{equation}
\Delta(x)=\Delta_0\cdot\tanh(x/(a\xi_0))
 \label{eq27F}
\end{equation}
The distortion to the homogeneous state
($\Delta(x)\equiv\Delta_0$) thus produced, is energetically
unfavorable. At large separation between the walls the resulting
energy loss is the sum of contributions originated near each
domain, $\sum E_s$. (Here $E_s$ is the one wall energy cost per
unit length).

Return now to the putative 1$^{st}$ order transition depicted on
Fig.1. To the homogeneous SDW/CDW phase corresponds the gain in
the Free Energy density:
\begin{equation}
\Delta F_{SDW}=-2\nu(\varepsilon_{F})\Delta_{0}^{2}
\label{Al1}\
\end{equation}
($\nu(\varepsilon_{F})=m/4\pi$  - DOS per single spin). But
deformations, $\eta_1$ and $\eta_2$, of the two initially
congruent FS's also decrease the energy of the normal (PM)state:
\begin{equation}
\Delta F_n=-2\nu(\varepsilon_{F})[\langle (v_F\eta_1)^2\rangle + \langle (v_F\eta_2)^2\rangle]
\label{Al2}\
\end{equation}

Hence, if for instance, the system is doped, the last mechanism
may offset the energy cost, $E_s>0$, of forming single DW owing to
the fact that there are locally gapless states inside the wall in
the form (\ref{eq27F}). Parametrizating doping, as before,
$v_F\eta_1=-v_F\eta_2=I$, one can write the local PM contribution
of Eq.(\ref{Al2})   in the form:
\begin{equation}
\Delta F_n(x) = -4\nu(\varepsilon_{F})\overline{\nu}(x/x_0)I^2
\label{Al3}\
\end{equation}
where $x_0$ is a scale for the domain width. Denote by
$F^\ast_{norm}(I)$ the integrated contribution from one single
domain (per unit wall length):
\begin{equation}
-F^\ast_n= -4\nu(\varepsilon_{F})I^2 \int \overline{\nu}(x/x_0) dx
\label{Al4}\
\end{equation}
The threshold value, $I_c$, is determined by:
\begin{equation}
E_s - F^\ast_n(I_c) =0
\label{Al5}
\end{equation}
At $I>I_c$ the spontaneous formation of domains will be
arrested by their interactions.

In principle, finding $E_s$, $\overline{\nu}(x/x_0)$ and the very
profile of $\Delta(x) \  (\ref{eq27F})$ can be reduced to the
self-consistent mean-field problem by solving Eqs.(\ref{eq24}) for
the band energy spectrum and the eigenfunctions in the periodic
potential, $\Delta(x+L)=\Delta(x)$. Near the wall where
$\Delta(x)$ has the form (\ref{eq27F}) Eqs.(\ref{eq26}) at $L\gg
x_0$ can be solved in terms of the hypergeometric functions.
However, the need to account the periodicity of
($\bar{u},\bar{v}$) at large separations between walls makes the
approach rather cumbersome.

With our main goal to attract attention to this new phenomenon we
choose the model \cite{GorMnaz} that allows the one-to-one mapping
on the FFLO problem. Therefore it is possible to avoid these
tiresome calculations by instead making the use of the numerical
results \cite{BurkRain}. Without staying on further details, we
obtained for $\overline{\nu}(x/\xi_0))\simeq 1.2
ch^{-1}(0.63x/\xi_0)$. Note that $\overline{\nu}(x/\xi_0))\sim 1$
for $x$ of the order of $\xi_0$, i.e. local DOS inside domain has
the same order of magnitude as in the PM normal phase. That is
also true in the periodic soliton lattice. With
$I_c=0.655\Delta_0$ we found from Eq.(\ref{Al5}) the soliton
energy
\begin{equation}
E_s=10.27\cdot\nu(\varepsilon_F) \xi_0 \Delta_0^2=0.26\cdot
p_F\Delta_0
 \label{Al7}
\end{equation}
(We accepted the BCS value $\xi_0=\hbar v_F/\pi\Delta_0$. The
Planck constant $\hbar$ is restored in the expressions for
$\xi_0$.)

With the soliton energy known, Eq.(\ref{Al7}), turn now to the
role of the anisotropy. Assume $\int\delta(\vec{p}_F)
dl_F\equiv0$, the equal numbers of electron and holes. Coming back
to the relation between $\tilde{E}(v_{F,x})$ and $E(v_{F,x},
\vec{p}_F)$ in Eqs.(\ref{eq24}) and (\ref{eq26}), one sees that
the energy spectrum of Eqs.(\ref{eq26}) for $(\tilde{u},
\tilde{v})$ is pertinent only to distorted homogeneous phase, when
negative contributions of the form of Eq.(\ref{Al3}) are absent.
The energy spectrum of Eqs.(\ref{eq26}), hence, contributes only
to calculations of the soliton energy $E_s$. Account of the
$\delta_{\vec{p}_F}$-terms,
$E(v_{F,x},p_x)=\tilde{E}(v_{F,x})-\delta_{\vec{p}_F}/2$,
decreases the cost of the single wall by filling up the momentum
states with $\delta_{\vec{p}_F}>0$:
\begin{equation}
E_{kin}=-2\int \frac{dl_F}{2\pi}|\delta_{\vec{p}_F}|/2
 \label{E}
\end{equation}
In the notations $t(p_{\parallel})\equiv\delta_{\vec{p}_F}/2$,
$p_{\parallel}=p_F\cos\varphi$, $t(p_{\parallel})$ now defines
the energy spectrum for carriers moving inside the wall.
For  the corresponding DOS $\nu_w(\varepsilon)$ (per single spin, per unit length) one has:
\begin{equation}
\nu_w(\varepsilon)=\int \delta(\varepsilon -t(p_{\parallel})) dp_{\parallel}/2\pi
 \label{F}
\end{equation}
By the order of magnitude $\nu_w(\varepsilon)\sim
(dt(p_{\parallel})/dp_{\parallel})^{-1}\sim p_F/t$; with $t\sim
\Delta_0$, $\nu_w(\varepsilon)\sim p_F/\Delta_0$. The DOS in
Eq.(\ref{F}) is concentrated inside the domain width $\sim \xi_0$.
Therefore the local DOS in the periodic soliton lattice with
$L\sim\xi_0$ is large, i.e., again of the same order as
$\nu(\varepsilon_{F})=m/4\pi$ the 2D DOS in PM state.

Thus $E_{kin}=E_s$ determines the threshold for onset of the
modulated SDW state at large enough geometric mismatch between the
two FS's. (The mechanism is akin to the one for forming periodic
domains in (TMTSF)$_2$PF$_6$ \cite{PF6new, GorGrig,BraGorSchr}.)

Up to this point we studied only the variation of the SDW state
with changes to the degree of nesting. Meanwhile, reducing the SDW
gap also opens way to emergence of a SC phase. Interactions
responsible for the 1$^{st}$ order magnetic and structural
transition seem to be stronger than the ones that lead to the SC
pairing. Indeed, in the stoichiometric phictides the former occurs
at higher temperatures. In the area of the phase diagram discussed
so far SC would develop on the background of the SP. Note that
when SC is still weak, its appearance may not fully remove remnant
DOS that, as we seen above, has by order of magnitude
the same value as in PM state.
This may explain the substantial residual density of states
towards 0 K, revealed via nuclear spin-lattice relaxation rate in
SC domains of SrFe$_2$As$_2$ \cite{Kitagawa} and via the finite
linear coefficient in the specific heat in SC state \cite{Hardy}.

The heterogeneous phase coexistence reported in experiments
\cite{Park} seem to realize the discussed scenario. Indeed, in
\cite{Park} the heterogeneous state first sets in below
$T_{SDW}\approx$70 K, i.e., well above $T_c\approx$32 K. While the
magnetic force microscope (MFM) images do not visualize a
periodicity in some special direction, this, actually, is not
expected. Below the 1$^{st}$ order transitions samples are
twinned. In addition,  unlike the strongly anisotropic
(TMTSF)$_2$PF$_6$ \cite{PF6new}, the spontaneous formation of
domains in pnictides can emerge along any arbitrary direction
leading to a pattern similar to the one seen in \cite{Park}. More
recently, the scanning force microscopy (SFM) measurements
\cite{Chuang} in CaFe$_{1.94}$Co$_{0.06}$As$_2$ revealed the
existence of striped electronic nanostructures with dimensions
$\sim$ 4 nm. Characteristic spatial scale for the superstructure
is $\xi_0 \sim 0.18\hbar v_F/T_{SDW}$ as obtained above. With the
typical $v_F\sim10^7$ cm/sec and $T_{SDW}\sim 100$ K for pnictides
it leads to few  nm. The $\mu$SR measurements \cite{Goko} were
capable to determine only the volume fractions of the coexisting
SDW and SC sub-phases. The hybrid SDW/SC structure found in
\cite{Kitagawa} may correspond to the case when $T_{SDW}$ is low
and close to $T_c$, but we have no results when SDW and SC compete
for existence on equal terms, i.e., under the SC dome near QCP in
Fig.1. Such a competition remains the subject of the great
interest (see, e.g., \cite{Vavilov, Khasanov}).

To conclude, we have shown that emergence of a heterogenic state
with large local DOS is an intrinsic property of iron pnictides at
low temperatures. Mechanisms of formation of a single DW near the
threshold were investigated unveiling the difference in the role
of doping and anisotropy. Although the simple model \cite{GorMnaz}
omits many details concerning the interactions and the energy
spectrum in real materials, it seems to confirm the view that
mesoscopic phase separation observed by means of NMR, $\mu$SR, MFM
and SFM, in reality is nothing but the formation of the system of
stripes.

\begin{acknowledgements}
The work of L.P.G. was supported by the NHMFL through NSF
cooperative agreement DMR-0654118 and the State of Florida, that
of  G.B.T. through the RFBR Grant N 10-02-01056.
\end{acknowledgements}


\begin{references}

\bibitem{Hosono1} Y. Kamihara \emph{et al.}, J. Am. Chem. Soc. 130, 3296 (2008)

\bibitem{Physica C} Special Issue on Superconductivity in Iron Pnictides,  Physica
C, 469, 313-674 (2009)

\bibitem{Park} J. T. Park  \emph{et al.}, Phys. Rev.
Lett., 102, 117006  (2009)

%\bibitem{Julien} M.-H. Julien \emph{et al.}, Europhys Lett. 87, 37001 (2009)

\bibitem{Kitagawa} K. Kitagawa \emph{et al.}, Phys. Rev. Lett. , 103, 257002 (2009)

%\bibitem{Nakai} Y. Nakai \emph{et al.}, Phys. Rev., B 81, 020503 (2009)



%\bibitem{Hosono2} Kenji Ishida, Yusuke Nakai, Hideo Hosono,  Journal of the Physical Society of Japan
%Vol. 78, No. 6, 2009, 062001

%\bibitem{Sad} M.V. Sadovskii, arXiv:0812.0302v1

\bibitem{Singh} D. J. Singh \emph{et al.}, arXiv:0810.2682

%\bibitem{Yi} M. Yi, D. H. Lu, J. G. Analytis, J.-H. Chu, S.-K. Mo, R.-H. He, M. Hashimoto, R. G. Moore, I. I. Mazin, D. J. Singh, Z. Hussain, I. R. Fisher, and Z.-X.Shen,
%Phys. Rev. B. 80, 174510 (2009); M. Yi, D. H. Lu, J. G. Analytis, J.-H. Chu, S.-K. Mo, R.-H. He, R. G. Moore, X. J. Zhou, G. F. Chen, J. L. Luo, N. L. Wang, Z. Hussain, D. J. Singh, I. R. Fisher, and Z.-X. Shen,
%Phys. Rev. B 80, 024515, (July 24,2009).

%\bibitem{Correlations} CORRELATIONS EFFECTS



\bibitem{Vavilov} M. G. Vavilov, A. V. Chubukov and A. B. Vorontsov,
arXiv:0912.3556v1; A. B. Vorontsov, M. G. Vavilov and A. V. Chubukov,
Phys. Rev. 79, 060508(R) (2009)

\bibitem{BarGor1} V. Barzykin and L.P. Gor'kov
Phys. Rev. B 79, 134510 (2009)

%\bibitem{Mazinrannij} Mazin, rannij

%\bibitem{Vorontsov} A.B. Vorontsov, M. G. Vavilov and A. V. Chubukov, Phys. Rev. 79, 060508(R) (2009)

\bibitem{KelKop} L. V. Keldysh and Yu. V. Kopaev, Sov. Phys. Solid State 6, 2219
(1965); A. N. Kozlov and L. A. Maksimov, Sov. Phys. JETP 22, 889
(1966).

\bibitem{LO} A. I. Larkin and Y. N. Ovchinnikov, Sov. Phys. JETP 20, 762 (1965).

\bibitem{FF} P. Fulde and R. A. Ferrell, Phys. Rev., 135, A550
(1964).

%\bibitem{Parker} D. Parker et al.,  Phys. Rev., B 80, 100508 (2009)

%\bibitem{BarGor2} V. Barzykin and L. P. Gor'kov, unpublished (2009).

%\bibitem{BarGorJETP} V. Barzykin and L. P. Gor'kov, JETP Lett. 88, 131 (2008).

\bibitem{PF6new} B. Salameh \emph{et al.}, Physica B: Condensed Matter, 404, 476
(2009);  N. Kang \emph{et al.}, Phys. Rev. B 81, 100509(R) (2010)

\bibitem{GorGrig} L. P. Gor'kov and P. D. Grigoriev, Europhys. Lett. 71, 425 (2005).

\bibitem{Rice} T.M. Rice, Phys. Rev. B 2, 3619,1970

\bibitem{GorMnaz} L. P. Gor'kov and T. T. Mnatzakanov, Sov. Phys. JETP, 36, 361, 1973

\bibitem{GorRus} L.P. Gor'kov and A. I. Russinov, Sov.Phys, JETP, 19, 922, 1964

\bibitem{Tesanovic} V. Cvetkovic and Z. Tesanovic, Phys. Rev. B 80, 024512 (2009)

%\bibitem{Laplace} Y. Laplace et al, Phys Rev B Rapid Com 80, 140501 (2009)

\bibitem{BurkRain} H. Burkhardt and D. Rainer, Ann. Phys. ( Berlin ) 3, 181(1994).

%\bibitem{AGD} A. A. Abrikosov, L. P. Gor'kov, and I. E. Dzyaloshinskii,
%Methods of Quantum Field Theory in Statistical
%Physics (Dover Publications, New York, 1977).

\bibitem{Eilenb} G. Eilenberger, Z. Phys. 214, 195 (1968)

\bibitem{ArtVolk} S.N. Artemenko and A.F. Volkov, in Charge Density Waves in
Solids, edited by L. Gor'kov and G. Gr\"{u}ner, Elsevier Science,
Amsterdam, 1989, Chap. 9.

\bibitem{LOnew} A. I. Larkin and Yu. N. Ovchinnikov, Sov. Phys. JETP 28,
1200, 1969

%\bibitem{BrazKir} S. A. Brazovskii and N. N. Kirova, Sov. Sci. Rev. A Phys., 5, 99
%(1984).
%\bibitem{BrazKir} S. Brazovskii and N. Kirova in Soviet Scientific Reviews,
%ed. by I. M. Khalatnikov (Harwood Academic, New York, 1984), Vol.
%5, p. 99

\bibitem{BraGorSchr} S. A. Brazovskii, L. P. Gor'kov, and J. R. Schrieffer, Phys. Scr. 25, 423 (1982)

%\bibitem{Takayama} H. Takayama et al PRB 21, 2388 (1980)

%\bibitem{PF6} D. Jerome, A. Mazaud, M. Ribault, and K. Bechgaard, J. Phys. (France) Lett. 41, L-95 (1980); R. L. Greene and E. M. Engler, Phys. Rev. Lett. 45, 1587 (1980).
%T. Vuletic et al., Eur. Phys. J. B 25, 319 (2002).

%\bibitem{ChubEremin} Chubukov, Eremin

%\bibitem{Mazin1} Mazin

%\bibitem{Parker} D. Parker et al.,  Phys. Rev., B 80, 100508 (2009)

\bibitem{Hardy} F. Hardy \emph{et al.}, Phys. Rev., B81, 060501(R) (2010)

\bibitem{Chuang}  T.-M. Chuang \emph{et al.}, Science 327, 181 (2010)

\bibitem{Goko} T. Goko \emph{et al.}, Phys. Rev., B80, 024508(2009)

\bibitem{Khasanov} M. Bendele \emph{et al.}, Phys. Rev. Lett. 104, 087003 (2010)

%\bibitem{Scalapino} S. Graser, T. A. Maier, P. J. Hirschfeld, D. J. Scalapino, New J. Phys. 11, 025016 (2009)



%\bibitem{Chubukov} Chubukov


%S.N. Artemenko and A.F. Volkov, in Charge Density Waves in
%Solids, edited by L. Gor'kov and G. Gru?ner ~Elsevier Science,
%Amsterdam, 1989!, Chap. 9.


%S. N. Artemenko  Phys. Rev. B 67, 125420 (2003)
%Model of charge-density-wave current conversion
%and phase-slip dynamics in mesoscopic samples








%\bibitem{Bud'ko} Bud'ko

%\bibitem{PF6} D. Jerome, A. Mazaud, M. Ribault, and K. Bechgaard, J. Phys. (France) Lett. 41, L-95 (1980); R. L. Greene and E. M. Engler, Phys. Rev. Lett. 45, 1587 (1980).
%T. Vuletic et al., Eur. Phys. J. B 25, 319 (2002).


%V. Stanev, J. Kang, and Z. Tesanovic, Phys. Rev. B 78, 184509
%(2008);

\end{references}
\end{document}